\pdfoutput=1
\documentclass[11pt,a4paper]{article}

\usepackage[margin=1in]{geometry}
\usepackage[T1]{fontenc}
\usepackage[utf8]{inputenc}
\usepackage{lmodern}
\usepackage{amsmath,amssymb,amsfonts}
\usepackage{graphicx}
\usepackage{booktabs}
\usepackage{float}
\usepackage{subcaption}
\usepackage{xcolor}
\usepackage{listings}
\usepackage{microtype}
\usepackage[hyphens]{url}

\usepackage[hidelinks]{hyperref}
\usepackage[nameinlink,capitalize,noabbrev]{cleveref}

\usepackage{authblk} 

\title{Money in Motion: Micro-Velocity and Usage of Ethereum’s Liquid Staking Tokens}

\author[1]{Benjamin Kraner\thanks{Corresponding author: \texttt{benjamin.kraner@uzh.ch}}}
\author[2]{Luca Pennella}
\author[3]{Nicol\`o Vallarano}
\author[3]{Claudio J. Tessone}

\affil[1]{\footnotesize{Blockchain \& Distributed Ledger Technologies Group, University of Zurich, Zurich, Switzerland}}
\affil[2]{\footnotesize{Department of Economics, Business, Mathematics and Statistics, University of Trieste, Trieste, Italy}}
\affil[3]{\footnotesize{UZH Blockchain Center, University of Zurich, Zurich, Switzerland}}

\date{} 

\begin{document}
\maketitle


\vspace{1em}

\begin{abstract}
We introduce a micro-velocity framework for analysing the on-chain circulation of Lido’s liquid-staking tokens, stETH, and its wrapped ERC-20 form, wstETH.
By reconstructing full transfer and share-based accounting histories, we compute address-level velocities and decompose them into behavioural components. Despite their growing importance, the micro-level monetary dynamics of LSTs remain largely unexplored. Our data reveal persistently high velocity for both tokens, reflecting intensive reuse within DeFi. Yet activity is highly concentrated: a small cohort of large addresses, likely institutional accounts, are responsible for most turnover, while the rest of the users remain largely passive. We also observe a gradual transition in user behavior, characterized by a shift toward wstETH, the non-rebasing variant of stETH. This shift appears to align with DeFi composability trends, as wstETH is more frequently deployed across protocols such as AAVE, Spark, Balancer, and SkyMoney. 

To make the study fully reproducible, we release (i) an open-source pipeline that indexes event logs and historical contract state, and (ii) two public datasets containing every Transfer and TransferShares record for stETH and wstETH through 2024-11-08.
This is the first large-scale empirical characterisation of liquid-staking token circulation. Our approach offers a scalable template for monitoring staking asset flows and provides new, open-access resources to the research community.
\end{abstract}


\section{Introduction}
The transition to Proof-of-Stake (PoS) in Ethereum \cite{buterinCombiningGHOSTCasper2020} has resulted in both expected and unexpected effects on the network. One notable outcome is the rise of third-party liquid staking providers \cite{scharnowski2025economics, gogol2024sok}. Staking in PoS protocols involves locking a specific amount of tokens (the stake) to participate in the consensus process. In Ethereum, stakers lock at least 32 ETH, enabling them to issue attestations in consensus epochs and participate in block proposal lotteries \cite{buterinCombiningGHOSTCasper2020,yan2024}. This stake acts as collateral to ensure honest behaviour; any dishonest activity may result in a penalty known as slashing. Additionally, staking requires participants to accept an opportunity cost. During the lock-up period, the staked tokens become illiquid, meaning they cannot be used in transactions or for other purposes. Liquid staking services address this limitation \cite{chitra2020stake}. 

Lido, the dominant provider of liquid staking services, allows users to convert ETH into \textit{stETH}, a liquid token. Holders of \textit{stETH} continue to earn staking rewards, minus a fee paid to the service provider. This fee compensates providers like Lido, the biggest LST protocol, for managing staking responsibilities, including tasks that, if mishandled, could result in penalties or fund losses \cite{cortes-goicoecheaAutopsyEthereumPostMerge2023}. In addition to \textit{stETH}, Lido also provides \textit{wstETH}, a wrapped version of the token designed to maintain a stable ratio relative to the underlying staked ETH. \textit{wstETH} is particularly suited for integration with decentralized finance (DeFi) protocols, as it is ERC-20 compliant and reflects staking rewards through an increasing exchange rate rather than token balance growth.

Liquid staking’s rapid adoption has raised concerns about possible unintended impacts on the PoS ecosystem. Critics argue that widespread use of liquid staking could lead to outcomes such as block cartelization, ambiguity in protocol governance, and increased coupling of capital risk with protocol risk \cite{RisksLSD}. Currently, PoS systems offer limited ways to mitigate the growth of liquid staking, with existing approaches focusing primarily on encouraging voluntary self-regulation by service providers \cite{RisksLSD}.

Liquid staking introduces a novel concept in the realm of digital assets, one that lacks a direct real-world comparison. It effectively allows money to multiply within users' digital wallets, a phenomenon largely absent from traditional financial systems\footnote{A useful comparison can be made with so-called \textit{``helicopter money''}, a monetary policy tool in which central banks distribute funds directly to households. However, in most advanced economies, such transfers are not the sole mechanism of money creation, nor are they directly tied to individual wealth. Another relevant analogy is a system of central bank digital currency (CBDC), where money exists solely in interest-bearing accounts managed by the central bank, unlike physical cash, which bears no yield.}. This unique characteristic raises important questions about the nature and classification of liquid staking tokens (LSTs) such as \textit{stETH} and its wrapped counterpart, \textit{wstETH}.

LSTs can be viewed as debt instruments: the staked tokens resemble the principal of a bond, the staking yield corresponds to interest payments, and the eventual return of staked tokens parallels principal repayment. This perspective aligns with their role as inflation-resistant stores of value. However, LSTs also exhibit unique characteristics, such as their use in DeFi applications as collateral or for yield farming, positioning them closer to programmable money.
Additionally, LSTs can be seen as derivative instruments, with their value tied to the underlying staked tokens. This interpretation underscores the complexity of LSTs and the challenges of fitting them into existing financial categories. The future trajectory of Ethereum's PoS ecosystem may depend on how these assets are perceived and used. If LSTs are primarily viewed as inflation-resistant stores of value, they may function like digital bonds. However, their liquidity and utility in transactions and DeFi applications could allow them to evolve into a new form of programmable money, which is secured against inflation from monetary expansion. 

Liquid staking has also introduced new risks to the Ethereum ecosystem. For example, the adoption of liquid staking services has raised concerns about centralisation. As of August 2022, Lido controlled a 30.1\,\% market share of total staked ETH, highlighting the significant influence these services can have on the network. Yet today, as of May 2025, the centralisation is only slightly attenuated, as Lido still has a 27\,\% market share of total staked ETH.

Moreover, the introduction of LSTs has increased the overall leverage in the crypto market. By enabling staked assets to be used as collateral, LSTs create new opportunities for yield farming but also introduce risk sources. The \textit{stETH} crisis in 2022 demonstrated how the perceived 1:1 peg between \textit{stETH} and ETH can break under market stress, leading to liquidity issues for overleveraged participants \cite{scharnowski2025economics}. As the Ethereum ecosystem continues to evolve post-merge, the role and impact of liquid staking will likely remain a critical area of focus. The balance between providing liquidity and maintaining network security, as well as the potential for new financial products and services built on LSTs, will shape the future of Ethereum's PoS system and the broader DeFi landscape.

In this work, we contribute to this debate by offering the first comprehensive analysis of LST usage and circulation dynamics through the lens of money micro velocity \cite{wang2003circulation, de2024ethereum, campajola2022microvelocity}, a granular measure of monetary activity at the individual account level. This approach enables us to detect patterns of financial behaviour that aggregate velocity metrics overlook, such as the concentration of activity among high net worth actors and the emergence of high-frequency intermediaries in an ostensibly decentralized ecosystem.

Our contributions are fourfold:
\begin{itemize}
    \item Methodological innovation: We adapt the micro velocity framework to rebasing tokens by reconstructing transfer histories in share denominated units, overcoming limitations in event log availability due to the late introduction of the TransferShares event in the Lido protocol.
    \item Empirical insights: Using on-chain data from December 2020 to November 2024, we compute and analyze both global and disaggregated micro velocities of \textit{stETH} and \textit{wstETH}, revealing a strong concentration of transactional activity in a small subset of large accounts.
    \item Comparative analysis: We contrast the behavior of rebasing \textit{stETH} and non-rebasing \textit{wstETH} LSTs, highlighting how ERC-20 compliance and integration with DeFi influence token usage and reutilization patterns.
    \item Open science and reproducibility: To support transparency and enable future research, we release all tools and datasets used in this study at \url{https://github.com/LucaPennella/money-in-motion-lsts}. This includes two open-source tools: one for extracting and indexing on-chain event logs, and another for querying historical contract state variables. We also provide two curated datasets: one containing \texttt{Transfer} events for \textit{wstETH}, and another with \texttt{TransferShares} records for \textit{stETH}. In addition, we publish the raw event logs and contract state variables used in our analysis.
\end{itemize}

The remainder of the paper is structured as follows. In the \textit{Related Work} section, we discuss prior research most relevant to our study. The \textit{Methods} section introduces the foundations of the micro velocity framework, provides an overview of the Lido platform and its on-chain architecture, the \textit{Data} section details the data collection process, and describes the preprocessing steps used to adapt micro velocity analysis to both \textit{stETH} and \textit{wstETH}. In the \textit{Results} section, we present the global micro velocity trends and their decomposition by user categories, followed by an analysis of balance dynamics and a characterization of the Lido ecosystem. Finally, the \textit{Discussion} section interprets the empirical findings and outlines directions for future research.

\section{Related Work}
The concept of \emph{micro velocity}, the turnover rate of money measured at the level of individual addresses rather than in aggregate, has recently enriched the economics of digital currencies \cite{wang2003circulation,de2024ethereum,campajola2022microvelocity, de_collibus_microvelocity_2025}.  Whereas traditional velocity treats all holders symmetrically, the micro perspective exposes behavioural diversity that is especially pronounced on blockchains.  Empirical work shows large cross-agent dispersion: wealthier accounts transact far more frequently and high-velocity \emph{intermediary entities} channel a disproportionate share of flows, hinting at latent centralisation in seemingly decentralised networks \cite{campajola2022microvelocity,de2024ethereum}.

Complementary research documents macro-structural asymmetries.  Makarov and Schoar \cite{makarov_blockchain_2021} reveal the dominance of a handful of players in Bitcoin, while network studies trace persistent hierarchical patterns and accumulative advantage across multiple chains \cite{de2024patterns,campajola2022evolution}.  Wealth inequality metrics such as the Gini coefficient broadly echo these findings \cite{sai_characterizing_2021}.  Governance-oriented work further shows how power concentrates through on-chain accountability mechanisms \cite{nabben_kelsie_accountability_2024}, and similar network tests have begun to quantify decentralisation in lending protocols such as Aave \cite{ao_is_2023}.

A third strand examines \emph{liquid staking protocols} (LSPs).  Gogol et al.\ provide the first systematic taxonomy and risk map of liquid-staking designs \cite{gogol2024sok,gogol_empirical_2024}; Scharnowski \cite{scharnowski2025economics} analyzes basis spreads in price discovery; and Tzinas \& Zindros \cite{tzinas_principalagent_2024} model the principal–agent tension between delegators and node operators.

Despite extensive work on native token circulation and the design of liquidity staking protocols, the intersection of staking, liquidity provision, and transactional dynamics at the agent level remains underexplored. Researchers have not yet systematically applied micro-velocity frameworks beyond native tokens, while LSP research rarely incorporates address-level behavioral analysis. In this paper, we bridge these gaps by extending the micro-velocity methodology to Lido’s rebasing (\textit{stETH}) and wrapped (\textit{wstETH}) tokens, uncovering distinctive patterns of liquidity concentration and transformation unique to liquid staking tokens (LSTs). In addition to velocity metrics, we incorporate address-level balance analysis, providing a deep dive into the temporal behavior of key actors. To support transparency and reproducibility, we also release curated open datasets and modular software tools that enable the community to replicate and extend our results. Together, these contributions offer a comprehensive empirical foundation for the study of liquid staking economies at scale.

\section{Methodology}
To analyze the circulation dynamics of LSTs, we adopt a micro-velocity framework that measures transactional activity at the level of individual accounts. This method allows us to move beyond aggregate metrics and identify heterogeneous behaviour among different user groups, capturing both active and passive usage patterns.

We first introduce the theoretical basis of micro velocity and how it can be adapted to account-based blockchains. Then, we outline its implementation for rebasing tokens like \textit{stETH}, including the handling of share-based accounting. Finally, we describe how this framework is extended to the wrapped, non-rebasing token \textit{wstETH}, enabling a comparative analysis of LST behavior within the Lido ecosystem.

\subsection{Micro Velocity}\label{sec:microvelocity}
Consider an ERC-20 token account $i$ endowed with a certain amount of tokens. Following \cite{de2024ethereum} we define the probability distribution of tokens holding times of agent $i$ at time $t$ as $P^t_i(\tau)$:
\begin{equation*}
    P_i^t(\tau) = \frac{w^t_i(\tau)}{M_i(t)}
\end{equation*}
where:
\begin{itemize}
    \item $w_i^t(\tau)$ is the total amount of tokens held by $i$ at time t with holding times $\tau$ and,
    \item $M_i(t)=\sum_\tau w_i^t(\tau)$ is the total amount of tokens held by $i$ at time $t$.
\end{itemize}
The Micro Velocity of agent $i$ is then defined as:
\begin{equation}
    V_i(t) = \sum_\tau \frac{1}{\tau} P_i^t(\tau) = \sum_\tau \frac{1}{\tau} \frac{w^t_i(\tau)}{M_i(t)}
\end{equation}
$V_i(t)$ dimension is $block^{-1}$ (i.e., measured per block). From $V_i$ we can derive the total velocity $M V$ of the token as:
\begin{equation}\label{eq:velocity}
    M(t) V(t) = \sum_i M_i V_i
\end{equation}

In order to compute $P_i^t(\tau)$ for an account-based blockchain (e.g., Ethereum), we adopt a Last-In-First-Out (LIFO) policy for token spending, where users first spend the most recently received funds; if those are insufficient, older balances with different ages are considered. 
This approach is economically meaningful, as it naturally distinguishes “liquid” money akin to M0 or M1 in traditional macroeconomics, which sits at the top of the wallet, from “illiquid” money, typically held as longer-term savings or investment. This LIFO assumption aligns with prior empirical work~\cite{de2024ethereum}, which also tested First-in-First-Out (FIFO) and random mixing policies and found negligible differences in velocity estimates. Intuitively, this reflects typical user behavior: everyday spending tends to use the most recently received or most accessible funds. For the numerical results, we relied on the Python package \texttt{MicroVelocityAnalyzer}\footnote{\url{https://github.com/fdecollibus/MicroVelocityAnalyzer/tree/parallelised_plus_bilance}}.

\subsection{Lido Platform Data Overview}
Lido is a liquid staking platform that enables users to stake their ETH in a liquid and fractionalized manner. Through Lido, users can acquire \textit{stETH} tokens by depositing ETH, which Lido stakes on their behalf. In exchange, \textit{stETH} or its wrapped version, \textit{wstETH} tokens are transferred to the user’s account, representing the staked ETH. Unlike standard staked ETH, which remains locked as collateral, these tokens retain liquidity, allowing users to transfer them freely. This feature enables users to invest amounts below the 32 ETH required to initiate a validator, without the operational responsibilities typically associated with validator duties.

As the staked ETH accumulates rewards, Lido distributes these rewards to \textit{stETH} holders through a process known as rebasing~\cite{gogol2024sok}. During rebasing, the \textit{stETH} balance in each user’s account is recalculated daily to reflect the earned rewards, effectively increasing the user's \textit{stETH} balance over time\footnote{  
Of consideration for the interested reader: the core Lido repository (current implementation): \url{https://github.com/lidofinance/lido-dao/tree/master} and a primer on Lido: \url{https://lido.fi/static/Lido:Ethereum-Liquid-Staking.pdf}.}.
In contrast, \textit{wstETH} does not rebase. Instead, it reflects staking rewards through a continuously increasing exchange rate relative to ETH. As a result, \textit{wstETH}  are fully interoperable with DeFi platforms where predictable token balances and ERC-20 compatibility are essential and can be bridged across multiple blockchains~\cite{gogol2024sok}. 

The data collected for this study includes the transfer records related to both \textit{stETH} and \textit{wstETH} tokens, capturing user transactions and the evolution of staking positions over time. Additionally, we extract the internal state of the \textit{stETH} smart contract, which is necessary to complement transfer information, particularly for accurately interpreting reward distributions, as will be discussed in the following sections.

\subsubsection{stETH Shares}\label{subsubsec:shares}

The \textit{stETH} contract employs an internal share-based accounting mechanism to track each user's stake in the Lido-controlled ETH pool. When a user acquires \textit{stETH} tokens, they effectively purchase shares in the Lido staking pool, representing the ETH they have staked. The conversion rate between \textit{stETH} and shares is dynamic and depends on the total ETH staked with Lido.

Two functions facilitate the conversion between \textit{stETH} tokens and shares:
\begin{itemize}
    \item \texttt{getSharesByPooledEth}: This function calculates the number of shares corresponding to a specified amount of \textit{stETH}, based on the formula:
    \begin{equation}\label{eq:steh2shares}
    \text{shares} = \frac{\text{stETH amount} \times \text{total shares}}{\text{total pooled ETH}}
    \end{equation}
    \item \texttt{getPooledEthByShares}: This function determines the amount of \textit{stETH} corresponding to a specific number of shares, following the inverse formula:
    \begin{equation}\label{eq:shares2steth}
    \text{stETH amount} = \frac{\text{shares} \times \text{total pooled ETH}}{\text{total shares}}
    \end{equation}
\end{itemize}

These functions ensure that the protocol can dynamically adjust share-to-token ratios, accurately reflecting changes in the total pooled ETH and thus ensuring that each user’s stake aligns with the current total pool distribution \footnote{Source: \url{https://github.com/lidofinance/lido-dao/blob/master/contracts/0.4.24/StETH.sol}}.

\subsubsection{Rebasing}
As rewards accrue from staked ETH, the \textit{stETH} balance for each user changes to reflect these earned rewards, without affecting the user's underlying shares. This process, known as \textit{rebasing}, recalculates each user's \textit{stETH} balance based on their share of the total pooled ETH. The calculation follows the formula from \cref{eq:shares2steth}\footnote{Source of the formula and example: \url{https://docs.lido.fi/contracts/lido\#rebase}}:

\[
\text{balanceOf(account)} = \frac{\text{shares[account]} \times \text{totalPooledEther}}{\text{totalShares}}
\]

where:
\begin{itemize}
    \item \textbf{shares}: A mapping of each user’s share count, updated with every Ether deposit, representing their fractional ownership in the staking pool.
    \item \textbf{totalShares}: The sum of all shares across accounts, used to proportionally distribute the pooled ETH.
    \item \textbf{totalPooledEther}: The total ETH held by the protocol, defined as the sum of \texttt{bufferedBalance}, \texttt{beaconBalance}, and \texttt{transientBalance}, where:
    \begin{itemize}
        \item \textbf{Buffered balance}: ETH stored in the contract, yet to be deposited.
        \item \textbf{Transient balance}: ETH submitted to the Deposit contract, pending visibility in the beacon chain state, defined as the difference between \texttt{DEPOSITED\allowbreak\_VALIDATORS\allowbreak\_POSITION} and \texttt{BEACON\allowbreak\_VALIDATORS\allowbreak\_POSITION} measured in ETH.
        \item \textbf{Beacon balance}: ETH held in validator accounts, reported by oracles and serving as the primary contributor to \textit{stETH} supply.
    \end{itemize}
\end{itemize}

Rebasing events are triggered by the \texttt{handleOracleReport} function, which emits the \texttt{TokenRebased} event to reflect updated values for \texttt{totalPooledEther} and \texttt{totalShares}. It is important to note that rebasing only affects the value of these two variables, which in turn affects the balance calculation of each account in the \textit{stETH} smart contract, but the rebasing doesn't iterate over all the accounts to update their balance, which would prove unfeasible and uneconomical.

\begin{table*}[!ht]
    \centering
    \caption{Summary of recorded events, highlighting that \texttt{TransferShares} events began later than \texttt{Transfer} events.}
    \label{tab:events-stats}
    \begin{tabular}{lllll}
    \toprule
    Token & Event & Number of Records & First Record Block & Last Record Block \\
    \midrule
    stETH & Transfer & 2,792,968 & 11,480,187 & 21,145,533 \\
    stETH & TransferShares & 2,519,615 & 14,860,275 & 21,145,533 \\
    wstETH & Transfer & 1,420,359 & 11,888,810 & 21,145,533 \\
    \bottomrule
    \end{tabular}
\end{table*}

\subsubsection{Minting Events and Initial Token Distribution}

For the scope of the present work, we are not interested in the intricate web of smart contracts that manage the deposit of ETH and minting of \textit{stETH}. What we care about is the trace (if any) that minting (and consequently new \textit{stETH} users' adoption) leaves on the \textit{stETH} token transfer records.

This process is observable through a standard on-chain event pattern: every token minting is recorded as a transfer from Ethereum zero address \footnote{Ethereum zero address: \texttt{0x0000000000000000000000000000000000000000}} to the minting user account, i.e. the user who's depositing ETH to buy \textit{stETH}.
This is consistent throughout all \textit{stETH} implementations in time.

\subsubsection{ETH Redemption and Shares Burning}
For the scope of our research, which is computing the micro velocity of \textit{stETH} shares, the burning of shares is not important, as burned shares are virtually motionless, meaning they are not to be transferred anymore, and as such their weight in the velocity computation becomes negligible as time passes. As of the Lido 2.0 implementation, deployed on May 12\textsuperscript{th}, 2023, following the Shappella upgrade, the burning address\footnote{Burning address: \texttt{0xD15a672319Cf0352560eE76d9e89eAB0889046D3}} has been introduced. Because \textit{stETH} token burning became relevant with the implementation of stake withdrawal (as \textit{stETH} tokens then became redeemable as well), it is reasonable to assume that we can ignore burning addresses previously set. Since the burning address only received tokens and never sent any (as expected for such an address), we do not consider it in the velocity calculation.

\subsubsection{Lido Events}
Events in Solidity serve as a bridge between smart contracts and external applications. When state changes occur on the blockchain, events provide a way to log and track these changes. The event data gets stored in transaction logs alongside the blocks, making them accessible to applications monitoring the blockchain. 

In this study, we focused our attention on two particular events, which are crucial for tracking the flow of \textit{stETH} and share allocations, as well as for understanding changes in pooled ETH over time, which directly impacts user balances and protocol dynamics:

\begin{itemize}
    \item \texttt{Transfer}: This standard ERC-20 event logs token transfers, enabling tracking of \textit{stETH} movements between accounts.
    \item \texttt{TransferShares}: This event logs share transfers specifically, complementing the ERC-20 standard by tracking the underlying share movements not captured by token-only events.
\end{itemize}

As of the current Lido implementation (version 2.0) these two events are emitted on the Ethereum chain together when a transaction in \textit{stETH} tokens is emitted. This reflects the atomic relation between \textit{stETH} shares and tokens.

\subsubsection{Lido Variables}
Constant state variables in Solidity serve as immutable values that are determined at compile time and stored directly in the contract's bytecode rather than in storage. This design choice makes them highly gas-efficient since reading these values doesn't require accessing blockchain storage. They're commonly used for fixed values like role identifiers, configuration parameters, or mathematical constants that won't change throughout the contract's lifetime. 
 
 In this study, we were interested in recovering the conversion rate between \textit{stETH} shares and tokens, which, as described in  \cref{subsubsec:shares}, required the collection of the following constant state variables of Lido smart contracts:
\begin{itemize}
    \item \texttt{lido.Lido.depositedValidators}
    \item \texttt{lido.Lido.beaconValidators}
    \item \texttt{lido.Lido.beaconBalance}
    \item \texttt{lido.Lido.bufferedEther}
    \item \texttt{lido.StETH.totalShares}
\end{itemize}

\subsubsection{wstETH Mechanics and Tracking} \label{subsubsec:wsteth}

\begin{figure}[htbp]
    \centering
    \includegraphics[width=1\linewidth]{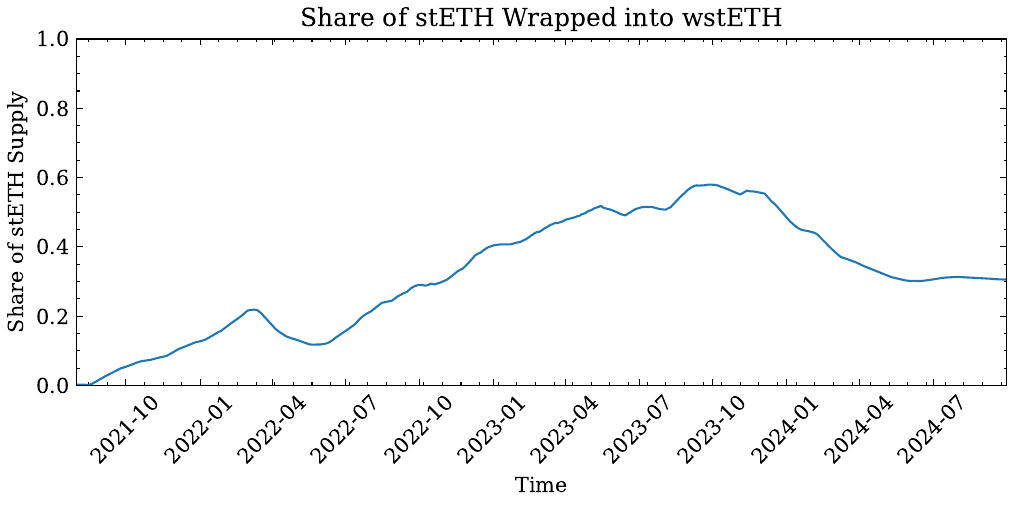}
    \caption{Evolution of the share of stETH wrapped into wstETH over time. The line represents the 30-day moving average of the proportion of total stETH supply held by the wstETH contract. This metric captures the relative demand for tokenized, non-rebasing stETH across the analyzed period.}
    \label{wstETH_share_over_time}
\end{figure}%

\textit{wstETH} is the wrapped, non-rebasing version of \textit{stETH}, designed to improve compatibility with decentralized finance (DeFi) protocols. Unlike \textit{stETH}, which periodically updates account balances to reflect accrued staking rewards (rebasing), \textit{wstETH} maintains fixed balances per account. Instead, it encodes staking rewards via an increasing exchange rate relative to \textit{ETH}, updated daily in sync with \textit{stETH}'s rebase events.

This structure makes \textit{wstETH} fully compliant with the ERC-20 standard and particularly suitable for smart contract integration, where deterministic token behavior is essential. As a result, tracking \textit{wstETH} on-chain is straightforward: the token emits standard \texttt{Transfer} events, and no share-based accounting is required.

In our analysis, we leverage the complete transfer history of \textit{wstETH} to compute micro velocity without needing to reconstruct internal state variables or account for rebasing. While this limits granularity compared to \textit{stETH}, it reflects realistic usage conditions and supports a comparative evaluation across LST designs. \Cref{wstETH_share_over_time} illustrates the proportion of total stETH supply wrapped into wstETH over time.

We now turn to the data sources and processing steps required to apply this framework to the Lido ecosystem.

\section{Data}
To implement the micro-velocity framework described in the previous section, we constructed a dataset that captures both the transactional and structural characteristics of Lido’s liquid staking tokens. This required overcoming several challenges stemming from the rebasing nature of \textit{stETH}, the evolving protocol design, and the need for consistent longitudinal tracking of token behavior at the address level.

In what follows, we describe the data sources, tools, and procedures used to collect, reconstruct, and process the relevant on-chain information. This includes the development of custom software to extract event logs and smart contract state variables, as well as the reconstruction of share-denominated transfer histories necessary for accurate velocity computation.

\subsection{Software Tools}
Many different tools exist to collect and analyze Ethereum token data, see \cite{zheng2020xblock}. Given the rebase nature of \textit{stETH}, the token is not fully ERC-20 compliant, which implies specific tools have to be developed. After a careful analysis of Lido smart contracts, we developed two software tools, the \texttt{ethereum-event-tracker} and the \texttt{ethereum-variable-tracker}, which were integrated in the data pipeline to produce the data we use in the present work.

The \texttt{ethereum-event-tracker} repository\footnote{
\url{https://gitlab.uzh.ch/bdlt/ethereum-event-tracker}
} was used to retrieve event records from Ethereum logs used in this study. \Cref{tab:events-stats} provides a summary of the recorded events, including record counts and the range of blocks in which these events were recorded. The \texttt{ethereum-variable-tracker} was used to retrieve the state of the constant state variables listed in the previous section for each block of Lido deployment.
It is immediate to observe in \cref{tab:events-stats}, that the number of \texttt{TransferShares} events is fewer than the number of \texttt{Transfer} events, which does not sound consistent with Lido implementation.
This follows the fact that the \texttt{TransferShares} event was introduced to the protocol as part of LIP-11\footnote{\url{https://github.com/lidofinance/lido-improvement-proposals/blob/develop/LIPS/lip-11.md}}, an update designed to support the protocol’s adjustments in anticipation of the Ethereum Merge\footnote{\url{https://ethereum.org/en/roadmap/merge/}}. 
As described in \cref{subsubsec:wsteth}, \textit{wstETH} emits standard ERC-20 \texttt{Transfer} events, enabling straightforward tracking

\subsection{Data Processing}

\begin{figure}[h]
    \centering
    \includegraphics[width=1\linewidth]{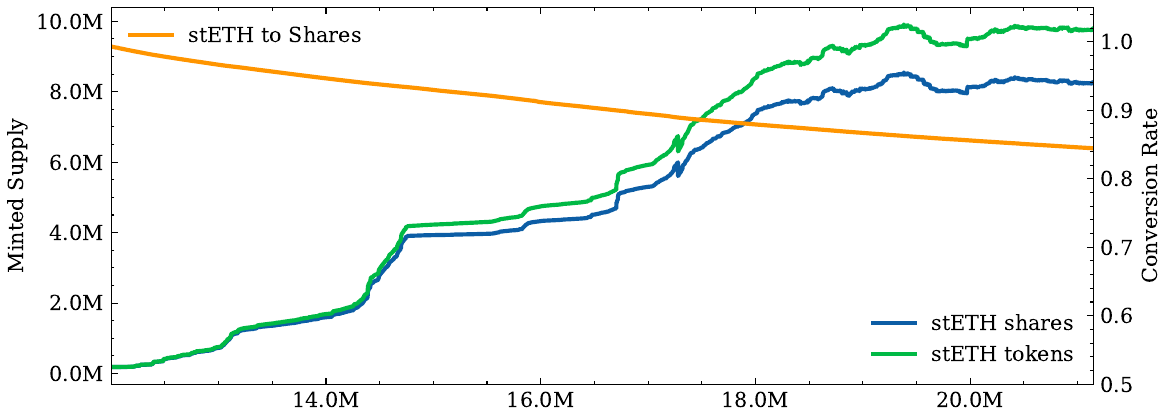}
    \caption{The left axis measures the the total amount of \textit{stETH} tokens (green line) and shares (blue line) in millions. The right axis measures the conversion rate from \textit{stETH} token to shares (orange line). On the x axis time is measured in Ethereum blocks.}
    \label{fig:supply}
\end{figure}

In order to compute the micro velocity of \textit{stETH} tokens, we had to account for the token's rebasing nature. 
An external observer who could only record transfers in \textit{stETH} tokens would not be able to infer the real balances of an account, due to the daily change in account balances denominated in \textit{stETH} tokens. Current methods to compute micro-velocity on Ethereum tokens rely on the collection of token transfers, and especially on the consistency of account balances denominated in the reference token. This makes it fundamental for our research to have access to the full record of \textit{stETH} transfers expressed in shares.

Because of the late introduction of the \texttt{TransferShares} event, the collection of share-denominated transfers is incomplete for the early period of \textit{stETH} existence. The same is not true for the \texttt{Transfer} events: as ERC-20 compatible events, these events were instantiated from the first deployment of the \textit{stETH} contract. To fill the gap, we recovered the state of the \textit{stETH} smart contract internal constants necessary to compute the tokens to share conversion rate and reconstruct the value in shares of each transfer originally recorded in \textit{stETH} tokens, using the formula described in \cref{subsubsec:shares}. This was done for each block of Lido deployment. By doing so, we were able to reconstruct a clean database of all \textit{stETH} transfers denominated in shares.

In \cref{fig:supply} we can observe the total supply of \textit{stETH} tokens and shares, as recovered from the constant state variable \texttt{lido.StETH.totalShares} from the \textit{stETH} smart contract. As the supply grows, the conversion rate diminishes, meaning that the value of a single share is increasing, as the total Ethereum pooled in Lido validators increases and, as such, the rewards they bear.
While this result is not novel per se, it stands to testify to the consistency of the data collection methodology. We plot the supply alongside the \textit{stETH} to Shares conversion rate, which is used to convert token transfers into shares transfers and complete the dataset.

We also compute micro velocity separately for \textit{wstETH}, the non-rebasing and ERC-20 compliant counterpart. Unlike \textit{stETH}, \textit{wstETH} maintains fixed balances and encodes staking rewards via an increasing exchange rate relative to ETH, thereby eliminating the need for share-based accounting. As such, the standard set of \texttt{Transfer} events is sufficient for computing micro velocity. While \textit{wstETH} does not offer the granularity of share-level tracking, its simplicity and compatibility with DeFi protocols make it an important asset for comparative analysis. The inclusion of \textit{wstETH} velocity allows us to capture a broader spectrum of user behaviour across Lido’s liquid staking products.

\section{Results}

\begin{figure}[h]
    \centering
    \includegraphics[width=1\linewidth]{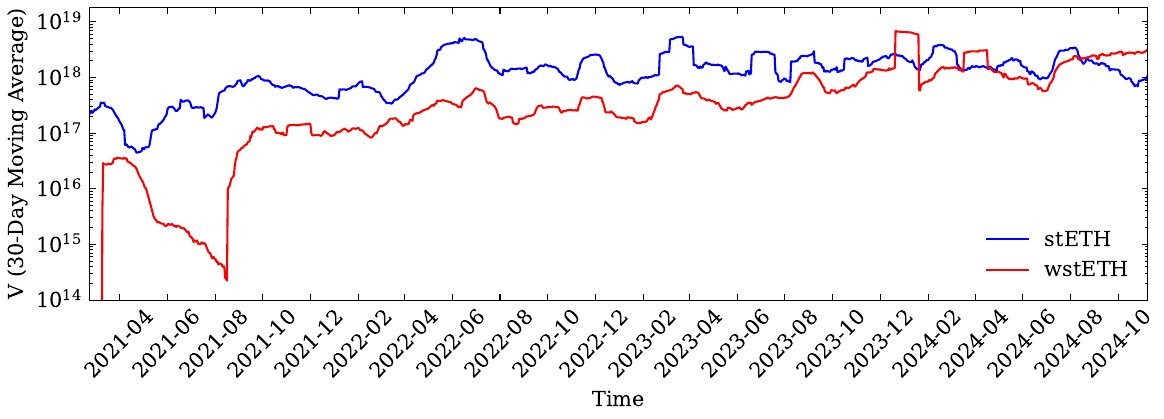}
    \caption{The global velocity $V(t)$ of \textit{stETH} and \textit{wstETH} tokens (see \cref{eq:velocity}, the dimension is $block^{-1}$, averaged weekly.)}
    \label{fig:velocity}
\end{figure}%

Following the methodology described in \cref{sec:microvelocity}, we calculate the individual \textit{stETH} and \textit{wstETH} velocities of all token accounts during the period under consideration, from block 11,480,187 (December 18, 2020) to block 21,145,533 (November 18, 2024). Using \cref{eq:velocity}, we derive the global velocity $V(t)$ for both assets, computed as the aggregate of individual micro velocities. The results are depicted in \cref{fig:velocity}, revealing three key features: consistently high values, a two-phase dynamic, and several distinct spikes.

By comparing the velocity we observe to the velocities reported on Proof-of-Work (PoW) and ERC-20 tokens in the literature \cite{campajola2022microvelocity,de2024ethereum}, we find that the global \textit{stETH}  and  \textit{wstETH} velocity is strikingly large. We find that both tokens exhibit significantly higher turnover. From a usage standpoint, we interpret this as evidence of Lido's success in achieving its mission: facilitating economic activity by unlocking the financial potential of staked tokens while maintaining Ethereum's Proof-of-Stake consensus and system security. Additionally, as interest-bearing tokens, \textit{stETH} and \textit{wstETH} provide an attractive alternative to ETH or other tokens as a store of value. They not only hedge against the effects of monetary expansion, thus appreciating in nominal terms, but also enable economic transactions.

\begin{table}[!htbp]
\centering
\caption{Summary of address categories by received \textit{stETH}. The \textit{Thresholds} column defines the lower and upper bounds used to assign accounts to categories.}
\label{tab:categories}
\begin{tabular}{lrrr}
\toprule
 & Thresholds (\textit{stETH}) & Total Received (\textit{stETH}) & Count \\
Category &  &  &  \\
\midrule
Whale &  $x\geq 10000$ & 141703559.627 & 1011 \\
Orca & $ 3000 \leq x < 10000$ & 6981034.142 & 1278 \\
Dolphin & $ 1000 \leq x < 3000$ & 4074250.128 & 2378 \\
Fish & $ 100 \leq x < 1000$ & 4547189.463 & 14441 \\
Shrimp & $ 10 \leq x < 100$ & 1636947.564 & 49982 \\
Krill & $ 1 \leq x < 10$ & 416861.963 & 113749 \\
Plankton & $ x < 1$ & 68469.071 & 307188 \\
\bottomrule
\end{tabular}
\end{table}

From \cref{fig:velocity}, we observe a two-phase trajectory: an initial growth phase followed by a mature phase. The growth phase occurred between the end of 2020 and the first half of 2022, during which velocity increased, reflecting greater adoption and the maturation of Lido's economic role. Notably, Lido's merge-ready protocol upgrade took place in May 2022\footnote{\url{https://research.lido.fi/t/announcement-merge-ready-protocol-service-pack/2184}}, coinciding with the Ethereum Merge, and introduced a more sophisticated version of the protocol. While this observation aligns with the natural growth in attention surrounding Lido and Ethereum's transition from PoW to PoS, it also validates our methodology, demonstrating its ability to capture significant on-chain economic events.

A significant spike in \textit{stETH} velocity is observed in March 2023. This aligns with user anticipation of the Shappella upgrade (April 2023), which introduced staking reward withdrawals, and the release of Lido 2.0 (June 2023). These events likely stimulated economic activity by unlocking previously illiquid capital.

Despite being a newer asset with a smaller user base, \textit{wstETH} shows a rapidly increasing velocity, converging with \textit{stETH} from early 2023 onward. This is particularly notable given the lower number of total addresses involved (\cref{tab:wsteth_categories}). We interpret this as a sign of higher per-token turnover, likely reflecting \textit{wstETH}’s enhanced suitability for DeFi applications due to its fixed-balance, ERC-20-compatible design. The asset’s structure encourages use in smart contracts and collateralized protocols, leading to intensive re-utilization. Notably, the global velocity of \textit{wstETH} increased above that of its base token \textit{stETH}. If this trend continues, differences in velocity may indicate a preferred role for staked Ether: \textit{wstETH} functioning as an inflation-resistant form of smart money, and \textit{stETH} serving more as a passive savings instrument.

To further investigate the source of the large velocity, we utilized the micro foundation of velocity and decomposed the global metric by contributor categories. Following the user taxonomy introduced by Lido\footnote{\url{https://blog.lido.fi/analysis-of-steth-user-behaviour-patterns/}}, we classified all \textit{stETH} accounts into seven categories based on the amount of \textit{stETH} received during the study period (December 2020 to November 2024). \Cref{tab:categories} and  \cref{tab:wsteth_categories} summarize these categories and their statistics for \textit{stETH} and \textit{wstETH}.

\begin{table}[ht]
\centering
\caption{Summary of address categories by received \textit{wstETH}. The \textit{Thresholds} column defines the lower and upper bounds used to assign accounts to categories.}
\label{tab:wsteth_categories}
\begin{tabular}{lrrr}
\toprule
 & Thresholds (\textit{wstETH}) & Total Received (\textit{wstETH}) & Count \\
Category &  &  &  \\
\midrule
Whale &  $x\geq 10000$ & 149285851.223 & 780 \\
Orca & $ 3000 \leq x < 10000$ & 4618320.295 & 850 \\
Dolphin & $ 1000 \leq x < 3000$ & 246.110 & 1401 \\
Fish & $ 100 \leq x < 1000$ & 2084328.725 & 6224 \\
Shrimp & $ 10 \leq x < 100$ &413353.129 & 11285 \\
Krill & $ 1 \leq x < 10$ & 54465.652 & 14063 \\
Plankton & $ x < 1$ &  8030.266 & 56518 \\
\bottomrule
\end{tabular}
\end{table}

For \textit{stETH}, whale accounts, that received at least 10,000 tokens, represent just 0.25\,\% of all addresses (1,011 out of more than 480,000) yet collectively received approximately 141.7 million \textit{stETH}, a dominant share of the total distribution. This confirms a well-documented trend in blockchain economies: a small number of large accounts command a disproportionate amount of monetary flow~\cite{de2024patterns, vallarano2020bitcoin, campajola2022evolution}. Mid-tier categories (e.g., Orcas and Dolphins) also hold substantial quantities, while the vast majority of accounts (Plankton, Krill, Shrimp) received comparatively minor volumes. 

A similar concentration is observed for \textit{wstETH}. Although whales are even fewer in number (780 addresses), they received nearly 149.3 million tokens, again constituting the majority of the total supply. Notably, the distribution for \textit{wstETH} is even more top-heavy: small holders (Plankton, Krill, Shrimp) are fewer in number than in the \textit{stETH} dataset and received a smaller aggregate amount (\textasciitilde 476k tokens vs. over 2.1 million for \textit{stETH}). This suggests a more targeted adoption of \textit{wstETH} among sophisticated or institutional users, possibly driven by its composability and compatibility with DeFi infrastructure.

Taken together, the distributions reported in \Cref{tab:categories} and \cref{tab:wsteth_categories} indicate that the usage of liquid staking tokens exhibits a marked degree of asymmetry. In both \textit{stETH} and \textit{wstETH}, a small number of high volume addresses, classified as whales, receive the majority of tokens, while the vast majority of accounts hold relatively modest amounts. This pattern suggests the coexistence of two distinct user profiles: on one hand, large entities that actively transact substantial volumes of LSTs, likely for integration within DeFi protocols; on the other, a broader base of smaller holders whose participation is limited to token acquisition, likely to passively benefit from staking rewards. 

\begin{figure}[htbp]
    \centering
    \begin{subfigure}[t]{0.9\linewidth}
        \centering
        \includegraphics[width=\linewidth]{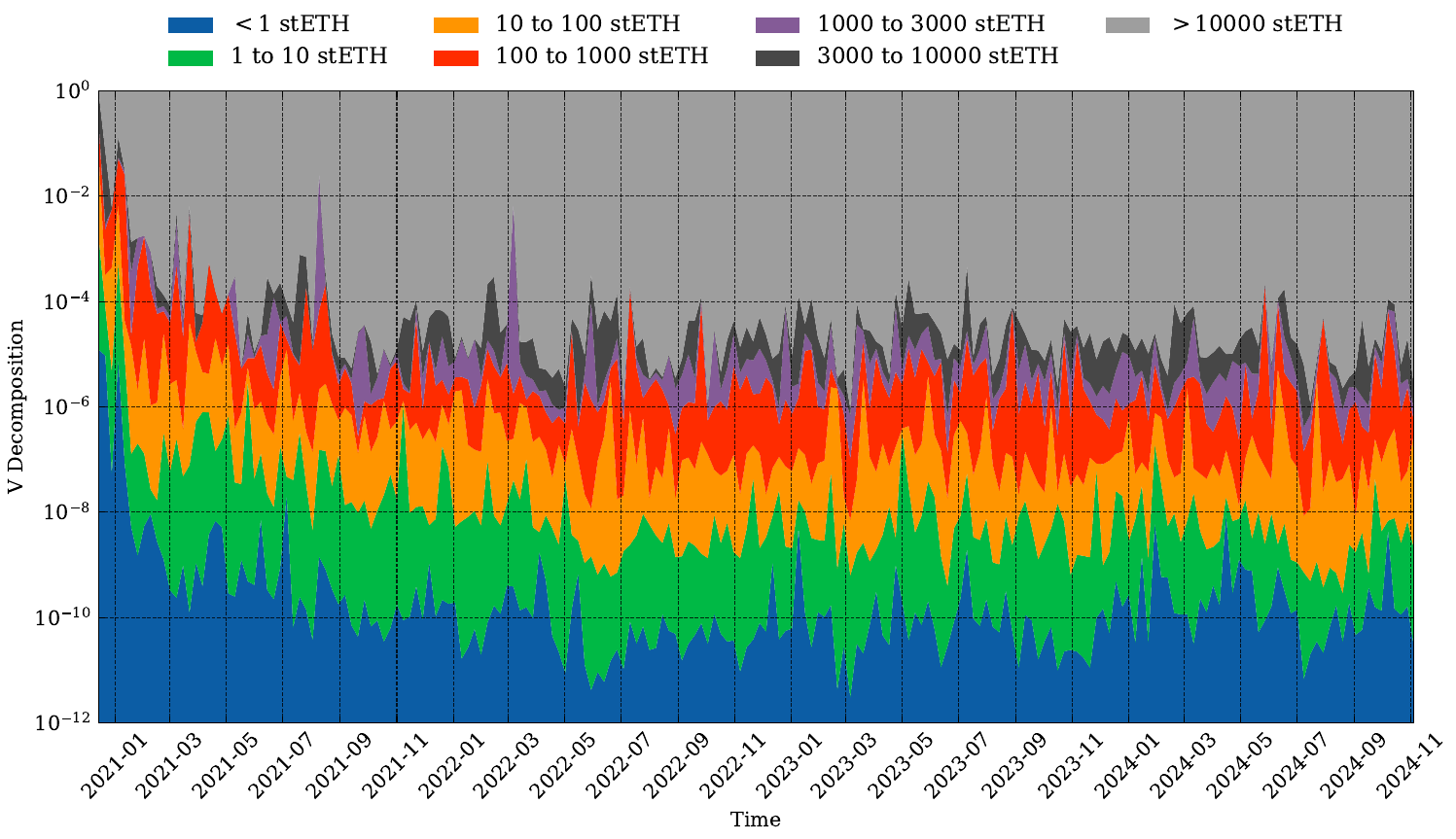}
        \caption{Micro velocity shares decomposition for \textit{stETH}, by categories defined in \cref{tab:categories}. Expressed in log scale.}
        \label{fig:steth-velocity-decomposition}
    \end{subfigure}

    \vspace{1em}

    \begin{subfigure}[t]{0.9\linewidth}
        \centering
        \includegraphics[width=\linewidth]{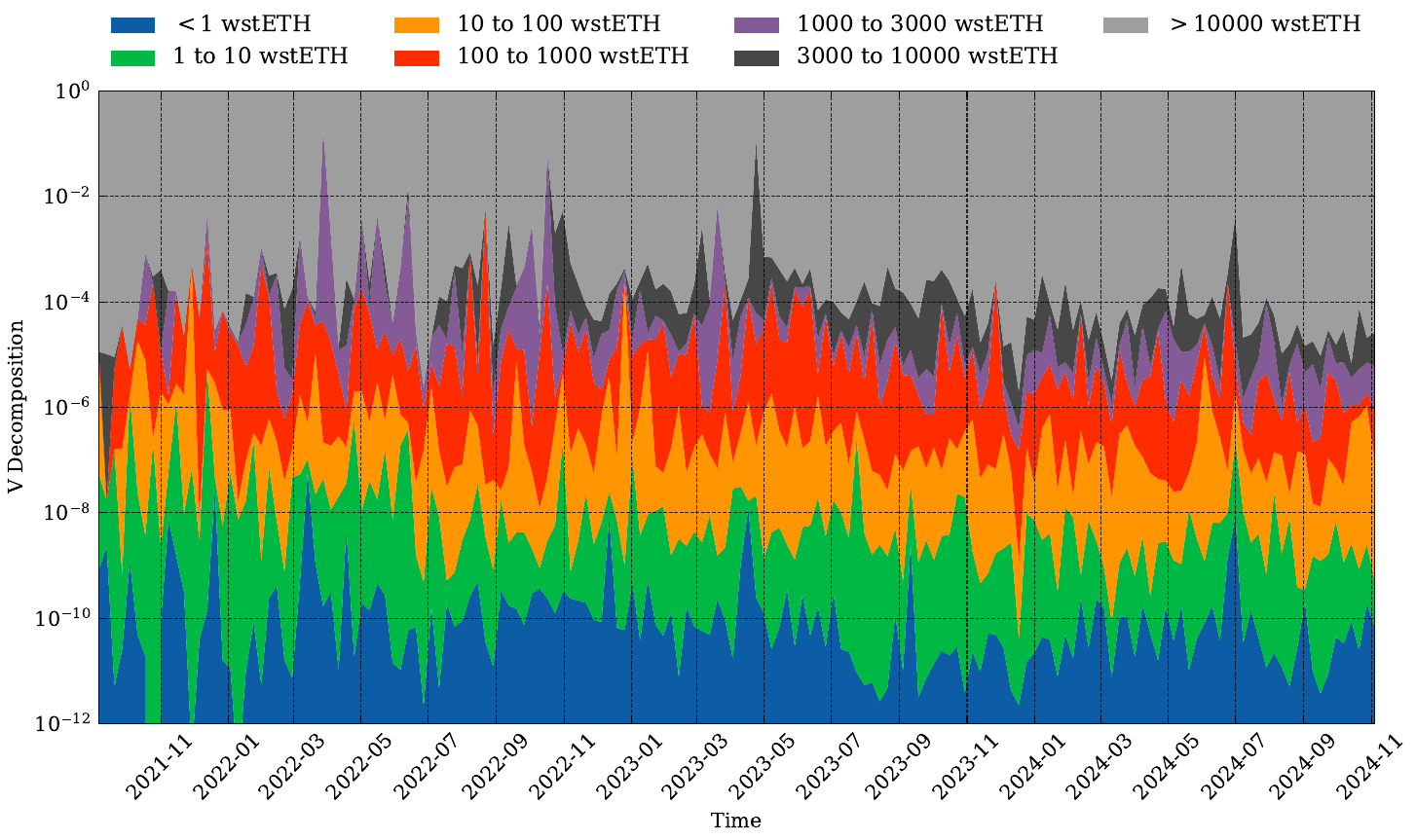}
        \caption{Micro velocity shares decomposition for \textit{wstETH}, by categories defined in \cref{tab:categories}. Expressed in log scale.}
        \label{fig:wsteth-velocity-decomposition}
    \end{subfigure}

    \caption{Comparison of micro velocity decomposition across address categories for \textit{stETH} and \textit{wstETH}. Both distributions are highly concentrated in Whale addresses.}
    \label{fig:velocity-comparison}
\end{figure}

When examining the decomposition of micro velocity by account category for both \textit{stETH} and \textit{wstETH} (\cref{fig:steth-velocity-decomposition} and \cref{fig:wsteth-velocity-decomposition}), a consistent concentration pattern is observed. In each case, whale accounts representing less than 1\,\% of the total account for the overwhelming majority of total velocity. For \textit{stETH}, whales contribute approximately 99\,\% of the aggregate velocity over the entire observation period. A comparable concentration is found for \textit{wstETH}, where high-volume accounts similarly dominate the transactional dynamics.

Outside the whale category, the remaining velocity contributions are substantially lower and display a more uniform distribution across mid and small volume groups. This stratification is more clearly visible in the log-scale representation of velocity shares, which reveals the limited role of lower-tier accounts in driving token circulation. These findings suggest that, while liquid staking tokens are widely distributed in terms of address count, the effective circulation is predominantly sustained by a narrow subset of large participants.

The observed concentration of velocity among whale accounts is likely influenced by the integration of liquid staking tokens into decentralized finance applications. Large accounts may correspond to actors such as DeFi protocols, liquidity providers, or institutional agents that engage in high-frequency and high-volume transactions, thereby contributing disproportionately to total velocity. This mechanism appears consistent across both \textit{stETH} and \textit{wstETH}, although the fixed-balance structure and DeFi-oriented design of \textit{wstETH} may further incentivize its use in automated financial operations. This analysis allows us to identify two distinct user typologies for Lido (and, by extension, for LSTs):

\begin{enumerate}
    \item A limited set of high-capacity accounts that receive and transact large volumes of liquid staking tokens, likely reflecting their integration into decentralized finance infrastructures.
    \item A broad base of smaller accounts that primarily hold LSTs in a passive manner, likely to access staking rewards without running a validator, and that exhibit low transactional activity.
\end{enumerate}

\begin{figure}[t]
    \centering
    \includegraphics[width=1\linewidth]{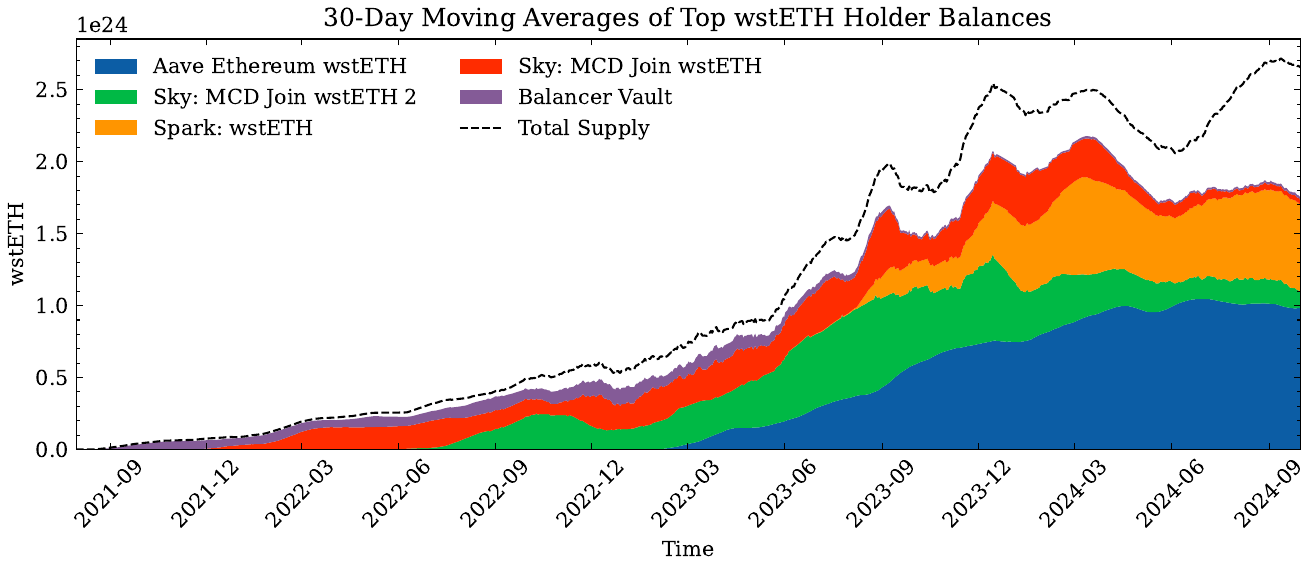}
    \caption{30-day moving averages of \textit{wstETH} balances for the top five holder addresses. These include major DeFi protocols such as AAVE, Spark, Balancer, and SkyMoney.}
    \label{fig:top-wsteth-holder-balances}
\end{figure}%

\begin{table}[!htbp]
\centering
\caption{Identified Ethereum Addresses and Associated Protocol Entities.}
\label{tab:address_labels}
\begin{tabular}{ll}
\toprule
\textbf{Address} & \textbf{Protocol Entities} \\
\midrule
\texttt{0x0b925ed163218f6662a35e0f0371ac234f9e9371} & Aave Ethereum wstETH \\
\texttt{0x12b54025c112aa61face2cdb7118740875a566e9} & Spark: wstETH \\
\texttt{0x248ccbf4864221fc0e840f29bb042ad5bfc89b5c} & SkyMoney: MCD Join wstETH 2 \\
\texttt{0x10cd5fbe1b404b7e19ef964b63939907bdaf42e2} & SkyMoney: MCD Join wstETH \\
\texttt{0xba12222222228d8ba445958a75a0704d566bf2c8}& Balancer Vault \\
\bottomrule
\end{tabular}
\end{table}

Beyond aggregate measures of micro-velocity, we further explore the behaviour of the largest token holders through account-level balance dynamics. \Cref{fig:top-wsteth-holder-balances} plots the 30-day moving averages of \textit{wstETH} balances for the five largest non-exchange holders, prominent DeFi actors such as Aave, Balancer, Spark and SkyMoney.

\Cref{tab:address_labels} lists these contracts. The Balancer V2 Vault is the upgradeable contract that escrow-holds every Balancer pool’s assets and executes swaps, joins, exits and external asset-management hooks; whenever a pool contains \textit{wstETH}, the Vault’s inventory rises accordingly, explaining its position as the single largest holder outside centralised exchanges. The Aave V3 \textit{wstETH} aToken and the homologous Spark Lend receipt token serve as interest-bearing ERC-20 wrappers. SkyMoney accepts \textit{wstETH} as collateral through two join adapters\footnote{\url{https://developers.sky.money/protocol/core/join/}}, whose dual-adapter architecture enables differentiated risk parameters. Collectively, these five contracts account for the overwhelming majority of \textit{wstETH} held outside centralised exchanges, illustrating the token’s deep integration into liquidity-provision, yield-generation and collateral-management workflows across the contemporary DeFi stack.

The time series reveals stable yet distinct trajectories, with entities like AAVE and Spark consistently maintaining large holdings, suggesting long-term integration of \textit{wstETH} into lending and collateral strategies.

It is noteworthy that an examination of the top five accounts already reveals a significant concentration of wstETH holdings, consistent with the broader concentration patterns observed in the velocity decomposition.

\section{Discussion \& Conclusion}
In this work, we provided a detailed analysis of the \textit{stETH} and \textit{wstETH} token transfer protocols, offering novel insights into the behaviour of Lido users by computing and decomposing the micro velocity of both liquid staking tokens (LSTs). By leveraging on-chain data, we captured the transactional activity of accounts across the entire lifespan of these assets, revealing both structural patterns and temporal dynamics.

Our findings show that both \textit{stETH} and \textit{wstETH} exhibit remarkably high global velocity, especially compared to previously studied Proof-of-Work assets and standard ERC-20 tokens. This supports the hypothesis that LSTs effectively unlock the liquidity of staked assets without undermining the security guarantees of Ethereum’s Proof-of-Stake consensus. As a result, these tokens circulate actively within the ecosystem while simultaneously accruing staking rewards. The increasing global velocity of \textit{wstETH} may serve as a proxy for its depth of integration and functional usability within DeFi applications, especially when compared to the velocity of its base token.

We also observe a growing proportion of \textit{stETH} that is wrapped into \textit{wstETH}, underscoring a demand for non-rebasing tokens, likely driven by their ERC-20 compliance and deterministic behaviour in smart contract environments. 

A key outcome of our decomposition is the identification of a highly asymmetric distribution of transactional activity. A small subset of accounts, classified as whales, is responsible for the vast majority of observed velocity in both \textit{stETH} and \textit{wstETH}, despite constituting less than 1\,\% of all users. These high-wealth actors likely represent institutional agents, DeFi protocols, and liquidity providers who integrate LSTs into automated or high-frequency financial strategies. Conversely, a large number of smaller accounts appear to use LSTs in a passive manner, primarily as a means of earning staking rewards without running a validator.

Further insights emerge when focusing on top holders' balance trajectories (\cref{fig:top-wsteth-holder-balances}). We observe persistent large-scale holdings by DeFi-native actors where a strong concentration is already evident among the top five addresses. This suggests a deliberate integration of the wrapped token into lending, collateralization, and composable DeFi protocols.

Taken together, these findings illustrate the emergence of a dual user base: on one hand, high-frequency DeFi participants actively leveraging LSTs within financial protocols; on the other, a broad population of low-activity holders passively engaging with staking infrastructure. This duality underscores the financial maturity and evolving specialization of LSTs in Ethereum's ecosystem.

To place our findings in proper context, the same micro-velocity lens must now be applied to other LSTs on Ethereum (e.g., \textit{cbETH}, \textit{rETH}) and to liquid-staking assets on external PoS chains. A cross-token panel will calibrate what “high” or “low” velocity really means and let us classify assets by behavioural archetype: passive store-of-value, actively rehypothecated collateral, bridge-hopping fuel, and so forth.  Because micro-velocity reacts at the transaction level, it can surface shifts in user behaviour long before coarse indicators such as raw volume or TVL respond, making it a candidate early-warning metric for researchers, protocol teams, and on-chain risk monitors.

For future work, it would be worthwhile to expand the analysis to \textit{yield farming} behaviours, whereby LSTs are recursively used as collateral to borrow and reinvest in additional staking positions. Such strategies, implemented through yield optimizers and money market protocols, may amplify both token velocity and systemic leverage. Identifying these patterns on-chain could help quantify endogenous risk and clarify their impact on the monetary dynamics of liquid staking ecosystems.

These results ultimately serve to reinforce the pivotal role of Lido within the context of Ethereum's staking and DeFi landscape. By providing liquid staking infrastructure that balances accessibility, composability, and economic utility, Lido has not only become the dominant LST provider but also a structural pillar of post-Merge Ethereum. It is imperative to comprehend the dynamics that are facilitated by metrics such as micro velocity in order to grasp the evolving monetary and systemic properties of Proof-of-Stake ecosystems.

\section*{Data Availability}
The datasets and tools used in this study are publicly available at \url{https://github.com/LucaPennella/money-in-motion-lsts}



\begin{thebibliography}{10}

\bibitem{ao_is_2023}
Ziqiao Ao, Lin~William Cong, Gergely Horvath, and Luyao Zhang.
\newblock Is decentralized finance actually decentralized? a social network analysis of the aave protocol on the ethereum blockchain.
\newblock {\em CoRR}, abs/2206.08401, 2022.
\newblock \href {https://arxiv.org/abs/2206.08401} {\path{arXiv:2206.08401}}, \href {https://doi.org/10.48550/arXiv.2206.08401} {\path{doi:10.48550/arXiv.2206.08401}}.

\bibitem{buterinCombiningGHOSTCasper2020}
Vitalik Buterin, Diego Hernandez, Thor Kamphefner, Khiem Pham, Zhi Qiao, Danny Ryan, Juhyeok Sin, Ying Wang, and Yan~X. Zhang.
\newblock Combining {GHOST} and casper.
\newblock {\em CoRR}, abs/2003.03052, 2020.
\newblock URL: \url{https://arxiv.org/abs/2003.03052}, \href {https://arxiv.org/abs/2003.03052} {\path{arXiv:2003.03052}}, \href {https://doi.org/10.48550/arXiv.2003.03052} {\path{doi:10.48550/arXiv.2003.03052}}.

\bibitem{campajola2022evolution}
Carlo Campajola, Raffaele Cristodaro, Francesco Maria~De Collibus, Tao Yan, Nicolò Vallarano, and Claudio~J. Tessone.
\newblock The evolution of centralisation on cryptocurrency platforms.
\newblock {\em CoRR}, abs/2206.05081, 2022.
\newblock \href {https://arxiv.org/abs/2206.05081} {\path{arXiv:2206.05081}}, \href {https://doi.org/10.48550/arXiv.2206.05081} {\path{doi:10.48550/arXiv.2206.05081}}.

\bibitem{campajola2022microvelocity}
Carlo Campajola, Marco D'Errico, and Claudio~J. Tessone.
\newblock Microvelocity: rethinking the velocity of money for digital currencies, 2023.
\newblock \href {https://arxiv.org/abs/2201.13416} {\path{arXiv:2201.13416}}.

\bibitem{chitra2020stake}
Tarun Chitra and Alex Evans.
\newblock Why stake when you can borrow?
\newblock {\em CoRR}, abs/2006.11156, 2020.
\newblock URL: \url{https://arxiv.org/abs/2006.11156}, \href {https://arxiv.org/abs/2006.11156} {\path{arXiv:2006.11156}}, \href {https://doi.org/10.48550/arXiv.2006.11156} {\path{doi:10.48550/arXiv.2006.11156}}.

\bibitem{de2024ethereum}
Francesco Maria~De Collibus.
\newblock {\em The ethereum ecosystem from a transaction network perspective}.
\newblock PhD thesis, University of Zurich, Z{\"u}rich, April 2024.
\newblock \href {https://doi.org/10.5167/uzh-260608} {\path{doi:10.5167/uzh-260608}}.

\bibitem{de_collibus_microvelocity_2025}
Francesco Maria~De Collibus, Carlo Campajola, and Claudio~J. Tessone.
\newblock The microvelocity of money in ethereum.
\newblock {\em {EPJ} Data Sci.}, 14(1):11, 2025.
\newblock \href {https://doi.org/10.1140/epjds/s13688-024-00518-6} {\path{doi:10.1140/epjds/s13688-024-00518-6}}.

\bibitem{cortes-goicoecheaAutopsyEthereumPostMerge2023}
Mikel Cortes{-}Goicoechea, Tarun Mohandas{-}Daryanani, Jose~Luis Mu{\~{n}}oz{-}Tapia, and Leonardo Bautista{-}Gomez.
\newblock Autopsy of ethereum's post-merge reward system.
\newblock In {\em {IEEE} International Conference on Blockchain and Cryptocurrency, {ICBC} 2023, Dubai, United Arab Emirates, May 1-5, 2023}, pages 1--9. {IEEE}, 2023.
\newblock \href {https://doi.org/10.1109/ICBC56567.2023.10174942} {\path{doi:10.1109/ICBC56567.2023.10174942}}.

\bibitem{de2024patterns}
Francesco~Maria De~Collibus, Carlo Campajola, Guido Caldarelli, and Claudio~J Tessone.
\newblock Patterns and centralisation in ethereum-based token transaction networks.
\newblock {\em Frontiers in Physics}, 12, 2024.
\newblock \href {https://doi.org/10.3389/fphy.2024.1305167} {\path{doi:10.3389/fphy.2024.1305167}}.

\bibitem{RisksLSD}
{Ethereum Foundation}.
\newblock The risks of lsd.
\newblock https://notes.ethereum.org/@djrtwo/risks-of-lsd, 2022.
\newblock Accessed: 2025-08-06.
\newblock URL: \url{https://notes.ethereum.org/@djrtwo/risks-of-lsd}.

\bibitem{gogol_empirical_2024}
Krzysztof Gogol, Benjamin Kraner, Malte Schlosser, Tao Yan, Claudio~J. Tessone, and Burkhard Stiller.
\newblock Empirical and theoretical analysis of liquid staking protocols.
\newblock {\em CoRR}, abs/2401.16353, 2024.
\newblock \href {https://arxiv.org/abs/2401.16353} {\path{arXiv:2401.16353}}, \href {https://doi.org/10.48550/arXiv.2401.16353} {\path{doi:10.48550/arXiv.2401.16353}}.

\bibitem{gogol2024sok}
Krzysztof Gogol, Yaron Velner, Benjamin Kraner, and Claudio~J. Tessone.
\newblock Sok: Liquid staking tokens (lsts) and emerging trends in restaking, 2024.
\newblock \href {https://arxiv.org/abs/2404.00644} {\path{arXiv:2404.00644}}, \href {https://doi.org/10.48550/arXiv.2404.00644} {\path{doi:10.48550/arXiv.2404.00644}}.

\bibitem{makarov_blockchain_2021}
Igor Makarov and Antoinette Schoar.
\newblock Blockchain analysis of the bitcoin market.
\newblock {\em SSRN Electronic Journal}, October 2021.
\newblock \href {https://doi.org/10.2139/ssrn.3942181} {\path{doi:10.2139/ssrn.3942181}}.

\bibitem{nabben_kelsie_accountability_2024}
Kelsie Nabben and Primavera~De Filippi.
\newblock Accountability protocols? on-chain dynamics in blockchain governance.
\newblock {\em Internet Policy Rev.}, 13(4):1--22, 2024.
\newblock \href {https://doi.org/10.14763/2024.4.1807} {\path{doi:10.14763/2024.4.1807}}.

\bibitem{sai_characterizing_2021}
Ashish~Rajendra Sai, Jim Buckley, and Andrew~Le Gear.
\newblock Characterizing wealth inequality in cryptocurrencies.
\newblock {\em Frontiers Blockchain}, 4:730122, 2021.
\newblock \href {https://doi.org/10.3389/fbloc.2021.730122} {\path{doi:10.3389/fbloc.2021.730122}}.

\bibitem{scharnowski2025economics}
Stefan Scharnowski and Hossein Jahanshahloo.
\newblock The economics of liquid staking derivatives: Basis determinants and price discovery.
\newblock {\em Journal of Futures Markets}, 45(2):91--117, 2025.
\newblock \href {https://doi.org/10.1002/fut.22556} {\path{doi:10.1002/fut.22556}}.

\bibitem{tzinas_principalagent_2024}
Apostolos Tzinas and Dionysis Zindros.
\newblock The principal-agent problem in liquid staking.
\newblock In Aleksander Essex, Shin'ichiro Matsuo, Oksana Kulyk, Lewis Gudgeon, Ariah Klages{-}Mundt, Daniel Perez, Sam Werner, Andrea Bracciali, and Geoff Goodell, editors, {\em Financial Cryptography and Data Security. {FC} 2023 International Workshops - Voting, CoDecFin, DeFi, WTSC, Bol, Bra{\v{c}}, Croatia, May 5, 2023, Revised Selected Papers}, volume 13953 of {\em Lecture Notes in Computer Science}, pages 456--469, Cham, 2023. Springer.
\newblock URL: \url{https://doi.org/10.1007/978-3-031-48806-1\_29}, \href {https://doi.org/10.1007/978-3-031-48806-1_29} {\path{doi:10.1007/978-3-031-48806-1_29}}.

\bibitem{vallarano2020bitcoin}
Nicolò Vallarano, Claudio~J. Tessone, and Tiziano Squartini.
\newblock Bitcoin transaction networks: An overview of recent results.
\newblock {\em Frontiers in Physics}, 8:286, 2020.
\newblock Published: December 3, 2020. Section: Physics of Networks.
\newblock \href {https://doi.org/10.3389/fphy.2020.00286} {\path{doi:10.3389/fphy.2020.00286}}.

\bibitem{wang2003circulation}
Yougui Wang, Ning Ding, and Li~Zhang.
\newblock The circulation of money and holding time distribution.
\newblock {\em Physica A: Statistical Mechanics and its Applications}, 324(3):665--677, 2003.
\newblock \href {https://doi.org/10.1016/S0378-4371(03)00074-8} {\path{doi:10.1016/S0378-4371(03)00074-8}}.

\bibitem{yan2024}
Tao Yan, Shengnan Li, Benjamin Kraner, Luyao Zhang, and Claudio~J. Tessone.
\newblock A data engineering framework for ethereum beacon chain rewards: From data collection to decentralization metrics.
\newblock {\em Scientific Data}, 12(1):519, 2025.
\newblock \href {https://doi.org/10.1038/s41597-025-04623-7} {\path{doi:10.1038/s41597-025-04623-7}}.

\bibitem{zheng2020xblock}
Peilin Zheng, Zibin Zheng, Jiajing Wu, and Hong{-}Ning Dai.
\newblock Xblock-eth: Extracting and exploring blockchain data from ethereum.
\newblock {\em {IEEE} Open J. Comput. Soc.}, 1:95--106, 2020.
\newblock \href {https://doi.org/10.1109/OJCS.2020.2990458} {\path{doi:10.1109/OJCS.2020.2990458}}.

\end{thebibliography}

\clearpage

\end{document}